\documentstyle[12pt]{article}
\begin{document}
\setlength{\baselineskip}{.7cm}
\renewcommand{\thefootnote}{\fnsymbol{footnote}}
\sloppy

\begin{center}
\centering{\bf FAULTS SELF-ORGANIZED BY REPEATED EARTHQUAKES IN A QUASI-STATIC
ANTIPLANE CRACK MODEL}
\end{center}

\begin{center}
\centering{Didier Sornette and Christian Vanneste\\
 Laboratoire de Physique de la Mati\`ere Condens\'ee,
CNRS URA 190\\ Universit\'e de Nice-Sophia Antipolis, B.P. 71\\ Parc
Valrose, 06108 Nice Cedex 2, France}
\end{center}

\renewcommand{\thefootnote}{\fnsymbol{footnote}}

{\bf Abstract :}
We study a 2D quasi-static discrete {\it crack} anti-plane model of a
tectonic plate
with long range elastic forces and quenched disorder. The plate is driven at
its
border and the load is transfered to all elements through elastic forces.
This model can be considered as belonging to the class of self-organized models
which may exhibit spontaneous
criticality, with four additional ingredients compared to sandpile models,
namely
quenched disorder, boundary driving, long range forces and fast time crack
rules.
In this ''crack'' model, as in the ''dislocation'' version previously
studied, we find
that the occurrence of repeated earthquakes organizes the activity on
well-defined
fault-like structures. In
contrast with the ''dislocation'' model, after a transient, the time evolution
becomes periodic with run-aways ending each cycle.
This stems from the ''crack'' stress transfer rule
preventing criticality to organize in favor of cyclic behavior.
For sufficiently large disorder and
weak stress drop, these large events are preceded by a complex space-time
history of
foreshock activity, characterized by a Gutenberg-Richter power law
distribution with
universal exponent $B=1 \pm 0.05$. This is similar to a power law
distribution of
small nucleating droplets before the nucleation of the macroscopic phase in a
first-order phase transition.  For large disorder and large stress drop, and
for
certain specific initial disorder configurations, the stress field becomes
frustrated
in fast time : out-of-plane deformations (thrust and normal faulting)
and/or a genuine
dynamics must be introduced to resolve this frustration.

\vskip 1cm
{\bf R\'esum\'e :}
Nous \'etudions un mod\`ele quasi-statique discret anti-plan de fissures
dans une plaque
tectonique en pr\'esence de forces \'elastiques \`a longue port\'ee et de
d\'esordre
gel\'e. La plaque est soumise sur ces bords \`a un taux de cisaillement
constant faible
qui est retransmis sur tous les \'el\'ements de la plaque par les interactions
\'elastiques. Ce mod\`ele peut \^etre vu comme un exemple de mod\`eles
auto-organis\'es qui peuvent donner lieu \`a une criticalit\'e spontan\'ee,
poss\'edant
les ingr\'edients additionnels suivants : d\'esordre gel\'e, for\c cage de
fronti\`ere,
transfert \`a longues distances et r\`egles d'\'evolution en ''temps
rapide'' de type
''fissure''. Dans ce mod\`ele de type ''fissure'', comme dans la version de
type
''dislocation'' pr\'ec\'edamment \'etudi\'ee, les tremblements de terre
r\'ep\'et\'es
s'organisent sur des structures de failles bien d\'efinies. A la diff\'erence
du mod\`ele de type ''dislocation'', on observe qu'apr\`es un transitoire
l'\'evolution
temporelle devient p\'eriodique, avec des \'ev\`enements catastrophiques
qui terminent
chaque cycle. Cela provient du transfert des contraintes \`a grandes
distances cr\'e\'ees par les fissures, qui emp\^eche la criticalit\'e de
s'organiser en
faveur d'un comportement cyclique. Pour des d\'esordres suffisamment grands et
des baisses de contraintes associ\'ees aux ruptures suffisamment faibles,
les grands
tremblements de terre sont pr\'ec\'ed\'es d'une histoire spatio-temporelle
complexe
d\'ecrivant l'activit\'e des pr\'ecurseurs, caract\'eris\'ee par une loi de
puissance de
type Gutenberg-Richter avec un exposant universel $B=1 \pm 0.05$. Ce
r\'esultat est
similaire \`a une distribution de petites ''gouttes'' de nucl\'eation
pr\'ec\'edant la
formation d'une phase macroscopique dans une transition du premier ordre.
Pour des forts
d\'esordres et de baisses de contraintes importantes lors des ruptures, et
pour des
r\'ealisations sp\'ecifiques du d\'esordre, le champ de contrainte devient
''frustr\'e''
pendant un tremblement de terre en ''temps rapide'' : des d\'eformations
hors-plan (failles
inverses et normales) et/ou une vrai dynamique doivent \^etre introduit
pour enlever cette
frustration.


\pagebreak

\section{Introduction}

In this paper, we continue our exploration of the hypothesis
according to which earthquake and fault characteristics can be
understood, at time scales of years and above,
only by using a global perspective,
treating on the same level the growth of faults by repeated earthquakes on one
hand and the localization of earthquakes on faults on the other hand.
A lot of studies have documented the self-similar structure of
fault patterns \cite{Kinga,Wu,Davy,Kagan,Kingb,Barton,Hanks,Scholza}. At
short time scales (typically less than or of the order of a century),
earthquakes
occur on these pre-existing set of faults and one can neglect the evolution of
the fault network to focus on the question of the role of the fault structure
in the observed earthquake phenomenology. On the other hand, faults
are evolving, nucleating, growing, branching, healing, dying eventually,
being screened by others faults \cite{Andrews}. This evolution occurs
as a result of the accumulation of deformations, accounted for by earthquakes
for a significant fraction (varying upon the location on earth). Therefore, the
fault geometry, which must be introduced to get a correct
description of earthquake occurrence, is not an arbitrary fractal, but results
from the accumulation of earthquakes and other more ductile modes of
deformations
which are themselves determined by the geometry. Our purpose here is to explore
further the implications of this ''hen and egg'' problem, within a simple
model.

To tackle this question, we have previously introduced a 2-D quasi-static
''dislocation'' model for the generation and organization of faults by repeated
earthquakes in an heterogeneous elastic plate driven at its border
\cite{Cowie,Miltenberger,Sornette}. The main results are the
spontaneous generation of fractal fault structures and the existence of a
well-defined
Gutenberg-Richter earthquake energy distribution $N(E) dE \sim E^{-(1+B)}
dE$, with
$B=0.3 \pm 0.05$ in 2D describing the earthquake population. The faults
have been
found to be globally optimal structures in the sense that they can be
mapped onto a
minimal interface problem, which in 2D corresponds to the random directed
polymer
problem \cite{Halpin}. More precisely, for small stress
drops, we have shown that a fault minimizes the sum of the random
thresholds of the
elements along it. This global minimization problem is achieved by the
spontaneous
organization of the medium, in which after a long ''learning'' transient
regime, the
deformation becomes localized on the optimal fault structure. In a sense,
the elastic
plate, endowed with its rules of rupture and stress redistribution, can be
viewed as an
analog computer which solves an optimization problem. Concretely, the
outcome of this
optimization correspondence is that the faults are self-affine with a roughness
exponent which is known exactly and equal to $2/3$.

In the present work, we study the
''crack'' version of the model : a constant ''dynamical'' stress is assumed to
characterize the ''fast time'' rupture, i.e. the stress on all broken elements
is fixed during a given earthquake event. In contrast, the
previous ''dislocation'' model \cite{Cowie,Miltenberger,Sornette}
 corresponds to imposing a slip to the ruptured
element, allowing it to be reloaded in fast time in the succession of ruptures
producing a complete earthquake. In the present ''crack'' model, an element
thus
fails only once in a given earthquake and stress enhancement occurs at the
crack
tips, growing, as usual, as the square root of the length of the evolving
earthquake. The motivation to study this variant of the initial model is
twofolds:
1) there is some debate in the litterature on the correct model (dislocation or
crack) to use for large earthquakes \cite{Scholzb,Roma,SS}; 2) in the context
of
self-organization, it is  important to assess the role of local rules in
the resulting
large-scale organization \cite{Gabrielov}.

We explore the various regimes as a function of initial disorder (on
the stress thresholds and elastic coefficients) and dynamical stress drop
amplitude. As
in the ''dislocation'' model, we find that the occurrence of repeated
earthquakes
organizes the activity on well-defined fault-like structures. The main
difference
with the ''dislocation'' model stems from the tendency for the model to
synchronize, i.e. to generate large periodic events, albeit of a complex
internal spatio-temporal structure. We interpret this behavior as resulting
from
the physics of coupled relaxation oscillators with threshold
\cite{Christensen,Sorn,Corral,Middleton,Bottani,Gil}.
For sufficiently
large disorder and weak stress drop, these large events are preceded by a
complex space-time history of foreshock activity, characterized by a
Gutenberg-Richter distribution with exponent $B=1 \pm 0.1$ which is
universal in the
sense that the exponent is found essentially the same for all systems explored.
For large disorder and large stress drop, for certain specific initial
disorder configuration, and after a long time, the model does not have a
solution
anymore: this corresponds to a situation where the stress field becomes
''frustrated''
and the anti-plane quasi-static crack modelling is no more defined. This
shows that the
quasi-static ''crack'' version of the model is not self-consistent and
additional types of deformations must be allowed to get rid of this
frustration. For instance, out-of-plane
deformations (thrust and inverse faulting) and/or the introduction of a genuine
dynamics can resolve this frustration. This breaking of self-consistency
is reminiscent of the breaking of unicity accompanying the appearence of
mechanical
instabilities for instance in elastic-plastic transition \cite{Lubliner}.

Ref.\cite{Lomnitz} has explored a variety of avalanches and epidemic
models which have the same type of stress enhancement transfer at the crack
tips.
In noiseless systems, they find, in agreement with us, periodic behavior and in
general large events of the size of the system.

\section{Description of the ''crack'' model}

The model is a direct extension to the crack case of the dislocation model
developed in
\cite{Cowie,Miltenberger,Sornette}. We consider
an elastic plate embedded in the (0x, 0y) plane and composed of plaquettes of
unit sizes paving the plane. The boundaries between the plaquettes constitute
the elementary fault segments. They are tilted at
$45$ degrees with respect to the $0x$ axis, ensuring a symmetric role for all
plaquette borders. A constant velocity boundary condition in the $z$
direction is
applied along the upper edge while the bottom edge is
kept fixed (both the upper and bottom edges are parallel to Ox). Due to
this externally
imposed deformation and the stress transfer due to elasticity, each
plaquette will
deform. Discretizing the mechanical problem, we attribute a single vertical
displacement
$w(x,y)$ along the direction $0z$ perpendicular to the plate, at the center
or node
$(x,y)$ of a plaquette. Each plaquette border is characterized by an
elastic constant
$g$ which may vary from element to element (quenched disorder on the
elastic coupling
coefficients). Only two components of stress are non-zero in this antiplane
model,
namely the stress $\sigma_{yz}(x,y)$ along $z$  applied on the border/fault
between the
plaquette centered on $(x,y)$ and the plaquette centered on $(x, y-1)$ given by
$\sigma_{yz}(x,y) = g [w(x,y) - w(x,y-1)] $ and the stress
$\sigma_{xz}(x,y)$ along
$z$ applied on the border between the plaquette  centered on $(x,y)$
and the plaquette centered on $(x-1, y)$ given by
$\sigma_{xz}(x,y) = g [w(x,y) - w(x-1,y)] $. Note that these expressions
are just the discretized version of Hooke's law for elasticity, expressed for
principal axis along Ox and Oy. For the present $45$ degrees tilted
lattice, the
formulas are deduced from those above by the standard rule of
transformation under
a rigid rotation. The elastic displacement $w(x,y)$ in the direction
$z$ normal to the lattice plane is solution of the discretized version of the
equilibrium elasticity equation $div \biggl( g(x,y) {\bf grad} w(x,y)
\biggl) = 0$.
Rupture occurs on a boundary between two
plaquettes when the stress applied on it reaches a threshold $\sigma_c$
which may
depend on the position (quenched disorder on the rupture thresholds)
\cite{Mohr}.
When an element breaks, the elastic strain in the element is
relaxed but the broken element suffers no change in its material properties
and it can
support stress again in the future. The stress field is assumed to obey the
equation
of mechanical equilibrium immediately after the rupture of an element.
The redistribution of elastic
stresses can bring other elements to rupture in a domino effect, creating model
earthquakes. What we
denominate as ''fast time'' is thus the succession of element ruptures
within an event,
during which the macroscopic load at the plate border does not increase
(''slow time''
is quenched during ''fast time''). This separation into these two time scales
is
intended to represent the difference between the fast dynamical rupture
which lasts
minutes at most compared to the tectonic loading which does not change over
this time
scale.

In our previous dislocation model, the nature of
the rupture on an element was simply characterized by the amplitude of the
slip,
chosen to be proportional to elastic deformation with a constant of
proportionality
$\beta$. This amounts to model a ruptured element as equivalent to a dipole
(antiplane
is scalar) whose strength is fixed until the element breaks again.
The total slip occurring on a
fault corresponds to the cumulative dipole amplitude on that fault. A large
earthquake
in the dislocation model can be viewed as a nonlinear rupture pulse
propagating in and
being multiply scattered by the heterogeneous medium, with the possibility
for an
element to rupture several times in fast time. In the present ''crack''
version, the strength of the dipole is not fixed in ''fast time'', but must be
reajusted at each rupture event in fast time during a given earthquake such
that
the dynamical stress $\sigma_{dyn}$, defined as $\sigma_{dyn} = (1-\beta)
\sigma_c$
where $\beta \sigma_c$ is the average stress drop, remains constant
and equal to a preassigned value on all rupture elements in this earthquake.
If $n$ elements have ruptured and a new element is
brought to rupture in fast time due to the stress redistribution induced by
these $n$
previous ruptures, the dipole strengths of the $n+1$ elements are
determined from a set
of $n+1$ equations as follows. Each dipole exerts a contribution to the
stress on all
other elements. The stress on any element is therefore the sum of the
background stress
prior to the earthquake plus the contribution of all the dipoles created in
the event.
These dipoles are then self-consistently determined such that the stress on
all the
ruptured elements in fast time is fixed and equal to the preassigned value.
Physically,
this models a situation where the faults remains ''open'' during the whole
duration of
the earthquake, at the opposite of the  dislocation model which corresponds
to an
instantaneous healing (closing) of the fault after each rupture.

In the crack model, there is another subtlety that was not present in
the dislocation model. Suppose two or more elements are brought above their
threshold
in fast time due to the stress redistribution. The correct physics would be
to solve
the elastodynamics equations which direct the evolution of these unstable
elements.
This is however too difficult to implement practically for an heterogeneous
system
with many interacting faults. Our quasi-static approach circumvents this
difficulty
at the price that one has to choose, rather arbitrarily, a rule for the
evolution
of the unstability. A priori, two rules can be introduced: 1) one
breaks them all simultaneously or 2) one ruptures only the element with the
largest
ratio (larger than one) of its stress to its threshold. We have checked
that these two rules
do not make any significant difference in the dislocation model. In the
crack model, only
the second rule has been explored in details in irreversible models of
rupture, whereas the
first rule may lead to un-ending ruptures. Our simulations have thus been
carried out with
this second rule. The elastic equations have been solved using a conjugate
gradient
technique with stopping criterion $10^{-20}$.

Most of our study will be carried out in the presence of quenched disorder in
the stress thresholds $\sigma_c$, which are drawn once for all
from a probability distribution $P_{\sigma}(\sigma_c)$ chosen uniform in
the interval
$[1-{\Delta \sigma \over 2}, 1+{\Delta \sigma \over 2}]$ with the value of
$\Delta \sigma$ between $0.1$ and $1.9$.

The basic difference between the dislocation and crack model is that stress
enhancement at the fault tip is much stronger in the latter, with a stress
growing as the square root of the fault length. As a consequence, the
nature of fault and slip organization depends on the system size $L$. An
earthquake
of size $L$ will generate a stress enhancement at its tip of
magnitude equal to $\beta \sigma_c \sqrt{L}$. Two cases appear: 1) if
$\beta \sigma_c \sqrt{L} < \Delta \sigma$, the amplitude of stress enhancement
generated by the dynamical evolution of the model is smaller than the quenched
heterogeneity. The latter thus dominates and we expect an organization similar
to that observed in the dislocation model where stress enhancement is small.
On the contrary, for $\beta \sigma_c \sqrt{L} > \Delta \sigma$,
sufficiently large
earthquakes will always create stress enhancements larger than the pre-existing
barriers. Beyond a characteristic nucleation size $L^*$ given by
$\beta \sigma_c \sqrt{L^*} \simeq \Delta \sigma$, earthquakes will not be
stopped
and will always break through the system (so-called ''run-away'' events).
In addition,
these large earthquakes will tend to smooth out the stress heterogeneity
along the
fault due to the condition of equal dynamical stress drop on the ruptured
elements.
These ingredients favor an approximate periodic state characterized by a
repetition of
large similar earthquakes
\cite{Christensen,Sorn,Corral,Middleton,Bottani,Gil}. Take for instance
$\beta = 0.1$.
Then, the run-away will be absent for systems sizes smaller than of the
order of
$100$. For such small stress drops, we have verified in particular
that the fault patterns selected in the
dislocation and the crack models have similar statistical properties in
this regime.
For larger stress drops or larger system sizes, there are two populations of
events : 1) the small ones similar to that occurring in the dislocation
model albeit
with a different distribution; 2) large earthquakes of size comparable to
the system
size.

\section{Threshold disorder}

Here, we wish to explore the crack regime. Our simulations have thus
been carried out on large systems $130$ by $130$. The influence of the
dynamical
stress $\sigma_{dyn}$ has been explored in the range $0 \leq \sigma_{dyn}
< 1-{\Delta \sigma \over 2}$, since otherwise unending ruptures occur (the
second
inequality expresses that the dynamical stress must obviously be smaller
than all static stress thresholds).

$\bullet$ $\Delta \sigma = 0.1$ (leading to $\sigma_{dyn} \leq 0.95$).

For $\sigma_{dyn}=0$ (large stress drop), after a short
transient, the dynamics is that of a perfect ''characteristic earthquake''
\cite{Schwartz}:  because of our tilted lattice
structure, a well-defined regular fault, made of two linear strands
oriented at $45$
degrees with respect to the  $0x$ axis and forming a $V$, is activated
regularly in a
perfect periodic fashion by a single great earthquake in which all elements
on the fault
break once in fast time with exactly the same slip. This regime corresponds
to a perfect
synchronization of all the threshold elements constituting the fault. The
Gutenberg-Richter
distribution is a Dirac function. The same is found for intermediate stress
drop
($\sigma_{dyn}=0.4$). For small stress drop ($\sigma_{dyn}=0.9$), we find
again a perfect
synchronization corresponding to the repetition of a single large identical
event. However,
the fault on which this event occurs is now rough, characterized by linear
strands
separated by rough portions. Its specific structure is however dependent
upon the specific
realization of the disorder on the thresholds. In summary, a small disorder
favors a very regular organization.

$\bullet$ $\Delta \sigma = 1$ (leading to $\sigma_{dyn} \leq 0.5$).
For $\sigma_{dyn}=0$, the behavior is very similar to the previous case
$\Delta \sigma = 0.1$, except that the fault is now rough all along its length.
For larger dynamical stress $\sigma_{dyn} \leq 0.2$ and $0.4$, the dynamics is
still periodic. However, a period contains a much more complex history than
just the succession, as documented until now, of a quiescent phase followed
by a single
great earthquake. In contrast, after a great earthquake, there is long
quiescent
phase, followed by the appearence of a diffuse foreshock activity spread
over the
plate. This diffuse activity is made of many small earthquakes. It
accelerates up
to the time where the great earthquake occurs on a fault. This fault is again
well-defined and is finally selected after a long transient. The distribution
of
earthquake energies contains two parts: a nice powerlaw distribution
$P(E) dE \sim E^{-(1+B)}$ for small
earthquakes $0.2 \leq E \leq 20$ (in the units where the elastic
coefficients are
all equal to $1$) with $B = 1 \pm 0.05$ and a Dirac peak at the energy of
the great
event (around $E \simeq 4000$). Notice the huge separation of energy
scales. This can
be rationalized using the nucleation argument outlined above. From the
expression
$\beta \sigma_c \sqrt{L^*} \simeq \Delta \sigma$, we get a nucleation scale
$L^*$ of the
order of $4$ elements. The energy being proportional to the square of the
length in
crack elasticity, this yields a characteristic maximum energy scale of $16$
not far from
the maximum energy observed in the power law distribution. In contrast, the
great
earthquake has a size of the order of one hundred and its release energy is
thus of
order $10^4$.

$\bullet$ $\Delta \sigma = 1.9$ (leading to $\sigma_{dyn} \leq 0.05$).
The progressive organization of the earthquake activity on a localized fault
network is shown in fig.1.
This case is similar to the previous one with the larger
$\sigma_{dyn}$. However, quantitatively, the rupture history during a period
separating the recurrence of two great earthquakes is even more characterized
by the appearence of a multitude of small earthquakes (see fig.2c). The
activity is always
localized on a well-defined fault structure (see fig.2a) which becomes
fixed at large times.
However, the fault is no more a linear object, but contains loops defining
internal
''micro-plates''. For the largest allowed dynamical stress drops (say
$\sigma_{dyn}=0.04$), we observe additional properties. After a great
earthquake,
there is, as usual, a quiescent time, followed by a progressing activation
of the
main fault by small earthquakes (fig.2c). The main fault is defined as the
locus of the
great earthquakes (see fig.2a). The activity of small earthquakes then
shifts progressively
away from the main fault to become delocalized in the bulk of the plate.
This activity
accelerates until the great earthquake occurs on the main fault. Again, we
observe
a huge separation of energy scales between the small events distributed
according to
a power law distribution for energies between $10^{-2}$ and a few $10^{-1}$
and the
great characteristic earthquake of energy of the order of $1500$. This is again
correctly explained by the nucleation argument. The exponent of the power law
distribution for small earthquakes is again $B = 1 \pm 0.05$ (see fig.2d)

Before the periodic regime organizes itself (see fig.2b) with its complex
spatio-temporal
structure of small earthquakes preceding the run-away, we witness a long
transient
aperiodic regime. The time duration of this transient is all the longer,
the larger
is the dynamical stress $\sigma_{dyn}$. For $\sigma_{dyn} = 0.04$ for
instance, the  transient is so long that one can measure with good accuracy the
distribution of earthquake energies in time windows sufficiently large so
that the
statistics is good, but sufficiently small so that the evolution of the
organization in
this transient regime is negligible. (Note however that the transient is
nevertheless
quite small compared to that observed in the dislocation model: the crack model
introduces a much stronger coupling via the crack stress redistribution rule,
favoring a faster and stronger organization). We again obtain a nice power law
for small earthquakes in the transient regime, with the same exponent $B$.
However,
the sizes of these ''small'' earthquakes in the transient regime are
typically one
order of magnitude larger than in the asymptotic periodic regime.

In summary, these numerical explorations show that the asymptotic dynamics is
always periodic, with however a more and more complex sequence of ruptures
within
a period, the larger are the disorder $\Delta \sigma$ and the dynamical
stress $\sigma_{dyn}$ (i.e. the smaller the stress drop). This can be
qualitatively
understood as an intermediate regime between the fully periodic regime, which
is
characteristic of low disorder and large stress drop, and the
self-organized critical
behavior observed for the dislocation model, which is controlled essentially by
large disorder and small stress drop \cite{Cowie,Miltenberger,Sornette}. We
have already underlined the analogy between
this problem and that of a set of interacting relaxation oscillators with
threshold. In this analogy, the stress drop parameter measures the coupling
strength
between elements, whereas the amplitude of the disorder $\Delta \sigma$
quantifies
the disorder in the natural frequencies of the individual elements. The analogy
predicts that synchronization, hence regular periodic behavior, will be the
stronger
the stronger the coupling and the weaker the disorder
\cite{Christensen,Sorn,Corral,Middleton,Bottani,Gil}.

Let us present a mean field toy version of the model, in the spirit of
\cite{Mirollo},
which allows one to understand the mechanism underlying the synchronization
process and
the appearence of  periodic states. For the purpose of clarity, consider an
homogeneous
system and the situation where the earthquake cycle is constituted of two
earthquakes,
involving respectively $N_1$ and $N_2 < N_1$ fault elements. The mean field
character of
the argument is to assume that, when an element reaches its threshold
$\sigma_c$, it
redistributes its stress to all the other $N = N_1 + N_2$ active elements,
each thus
getting a stress increment equal to $\alpha {\sigma_c \over N}$ where we
allow for an
arbitrary {\it positive} coupling strength $\alpha$. For simplicity of the
argument,
suppose also that the stress of the ruptured element is put to zero.
Suppose that we
start  reasoning at the time where the stress on the elements of fault $1$
is $\sigma_1 >
\sigma_2$, where $\sigma_2$ is the stress on the elements of fault $2$.
Earthquake $1$
will occur first, when  all $N_1$ elements are at their $\sigma_c$ in fast
time. At that
time, the stress on the elements of fault $2$ is $\sigma_2 + \sigma_c -
\sigma_1$, due to
the uniform tectonic loading. Just after event $1$, the stress on the
elements of $1$ is
zero by our rules while the stress on the fault $2$ is
$\sigma_2 + \sigma_c - \sigma_1 + {N_1 \alpha \sigma_c \over N}$. Fault $2$
will then
reach itself $\sigma_c$ at which time the stress on fault $1$ is
$\sigma_1 - \sigma_2 - {N_1 \alpha \sigma_c \over N}$. After earthquake $2$,
the stress on fault $2$ is zero by definition while the stress transfer
loads fault
$1$ to the level $\sigma_1 - \sigma_2 - {(N_1-N_2) \alpha \sigma_c \over
N}$. This last
result shows that the stress difference, which was initially $\sigma_1 -
\sigma_2$
has decreased during an earthquake cycle. The largest earthquake ($1$ in
this example)
is an absorbing state. This argument summarized here for an
homogeneous system works also for heterogeneous systems, if the disorder is not
too large \cite{Bottani}.

\section{Elasticity and threshold disorder}

We explore in this section the possibility to destroy the periodic
behavior and kill the great earthquake, by introducing disorder also
in the elastic coefficients of the elements and by varying the nature of
the disorder, for instance by allowing for the existence of very strong
and rigid elements (described by a powerlaw  distribution of thresholds
and/or elastic coefficients). This comes about because, the stronger
the disorder, the stronger will be the barriers to stop the run-away.

With respect to the fault structure and the time history, the addition of
disorder on
the elastic coefficient, when not too large, is tantamount to increasing
the threshold
disorder with no elastic disorder: we observe a periodic behavior, after a
transient
which is significantly larger than previously, all parameters being the
same otherwise.
When measurable, the distribution of small earthquakes is a power law with
an exponent
$B$ always equal to $1 \pm 0.05$. Fig.3 presents such a simulation with
$\Delta \sigma = 1.9$, a disorder on the elastic coefficients defined as for
the
threshold with a flat distribution of elastic coefficient and a width equal
to $\Delta g =
1$ and $\sigma_{dyn} = 0.04$.

For large threshold disorder
$\Delta \sigma = 1.9$, we find the novel feature that, in some system
realizations, the
time history does not seem to become periodic, however long we wait. A
global stationary
regime seems to emerge nevertheless, with an elastic energy stored in
the plate which fluctuates around a well-defined average. The fault
structure becomes
very rough with overhangs (this is allowed in the scalar anti-plane elastic
model used
here but would be unphysical in in-plane stress or strain models since
compressive and
extensive stress would accumulate without limit in the region of the overhang,
according to a ''hook'' effect \cite{Sordid}. Furthermore, the run-away does
not
exist anymore and is replaced by a continuous distribution of earthquake of
all sizes.

We have also explored different disorders, for instance a powerlaw
distribution of
rupture thresholds $P_{sigma}(\sigma_c) \sim \sigma_c^{-(1+\mu_{\sigma})}$ for
$\sigma_c \geq 1$, with
$\mu_{\sigma} = 0.5$ and $3$. The first case corresponds to a very broad
distribution (the average is mathematically divergent) with a significant
fraction
of the bonds having a very large threshold. Depending up the specific
realization, we find a similar phenomenology as before. For instance, for
$\sigma_{dyn} = 0$, a single great run-away punctuated the periodic dynamics.
However, the fault becomes very complex, with many branches and loops.
Correlatively,
the number of elements rupturing in the
great earthquake is about three times the system width.

For other disorder realizations with the powerlaw distribution of rupture
thresholds, in
presence or absence of elastic disorder, we have found a completely new
effect, that
we identify as a ''frustration'' which destroys the self-consistency of the
quasi-static crack model. This also occurs for the previous bounded
distribution in the
presence of quenched disorder in the elastic coefficients. The phenomenon
is best
illustration by examination of figure 4. It shows a small part of the
stress field
during an earthquake in fast time in a network with a power law distribution of
thresholds with $\mu_{\sigma}=0.5$ and $\sigma_{dyn} = 0.5$. The fault
segments which
have ruptured are indicated by an arrow with a thick line. These ruptured
elements all
carry a stress whose absolute value is $\sigma_{dyn} = 0.5$. The arrows
indicate the
sign of the stress: downward (resp. upward) arrows correspond to positive
(resp.
negative) stresses (recall that all the stresses are along the $Oz$ axis)
\cite{electric}. The stress carried by each fault element is indicated on
the arrow. The
rupture threshold for each element is written on the other side of the bond.
The
mechanical equilibrium translates into the condition that the sum of
stresses carried
by the fault segments attached to the same node be zero, easily verified in
this example. The self-consistency of the model is obeyed when the mechanical
equilibrium is such that all stresses are smaller than the corresponding
threshold.
When this is not the case, a rupture occurs relaxing the stress on the
rupture element
to $\sigma_{dyn}$. However, there are ''frustrated'' situations in which
this is not
possible. To see this, examine the element carrying a stress equal to $1.5$
whose
threshold is $\sigma_c = 1.495$.
By the definition of the model, since its stress is
larger than its threshold, it must rupture and its stress must thus be
lowered to
$\sigma_{dyn} = 0.5$. When this occurs, the mechanical equilibrium  is
broken, because
the three other fault elements connected to the same node have all their
stresses imposed
to be equal to the dynamical stress. The model looses its self-consistency
due to an
over-determination. In general, it is straightforward to realize that this
''frustrated''
state occurs when three fault elements connected to the same node have
ruptured. As a
consequence, their stresses is imposed to be equal to $\sigma_{dyn}$ and
from the rule
of quasi-static equilibrium, the stress on the remaining element connected
to the same
node is completely determined. If it happens that its threshold be less
than this
stress, the frustration appears and the model collapses.

This frustration is reminiscent of the concept of frustration introduced
in the physics of general disordered systems \cite{Toulouse}. In general,
frustration arises in a system whose interactions compete or conflict in
such a way
that not all constraints on the system can be simultaneously satisfied.
This is what happens in our model with the conflict between the stress
conservation and the dynamical stress which may produce frustration.
The usual outcome of frustation in the presence of disorder is the existence
of,
not a single well-defined stable equilibrium state, but rather to a very large
number of disordered equilibrium states of equivalent energies
\cite{Mezard}. The number
of these minimum states usually increases exponentially with the number of
degrees of
freedom (spins). The energy landscape is extremely complicated with a
hierarchy of
barriers of increasing sizes separating the minimum states. In other words,
in order to
go from one minimum state to another, an energy barrier must be passed.
This hierarchical
structure produces novel behaviors (long time relaxations, breakdown of
ergodicity, etc.)
and important fluctuations. In our model, mechanical equilibrium is not
more possible in
the presence of frustration, as we have shown. It is tempting to extrapole
from this
analogy and suggest that the state of stress chosen by the plate, in an
extended model
allowing for a genuine dynamics, could have a multi-valley structure
similar to that
obtained in spinglasses for instance. This could be an important underlying
ingredient at
the origin  of the observed spatio-temporal complexity.
In this spirit, we have recently proposed that the mechanics of coupled
blocks and
especially the coupling between rotations exhibit the frustration property
\cite{Frust}.
We believe that this is an ubiquitous and key property at the basis of the
complexity
observed in fault and earthquake organization.

The present crack model must be enriched to get rid of the
overdetermination. The
conceptually simplest solution is to introduce a genuine dynamics allowing a
transient breakdown of static equilibrium. For instance, elastodynamics allow
for an unbalance of the stress, corresponding to the generation and radiation
of
elastic waves. Physically, this describes the fact that the stored elastic
energy is
converted into two types of dissipation: 1) frictional heating, that we
take into
account by our rupture criterion and 2) elastic wave radiation that we
neglect. A
model in which the dissipation is not completely converted into friction,
i.e. the
stress is not immediately put to its asymptotic dynamical value, would cure the
observed frustration and over-determination.
Another alternative, within quasi-static mechanics, is to introduce
other types of deformations, such as thrust or normal faulting.
Algorithmically, this
will lead to allow a fraction of the stress to be removed by going out of
plane,
thereby removing the constraint of zero antiplane stress at all nodes.
However, we
still expect frustation to occur for certain realizations due to the
competition
between static mechanical equilibrium and constant dynamical stress.

It is interesting to note that there is a particular value of the dynamical
stress
drop for which the frustration never occurs, namely for
$\sigma_{dyn}=0$. In this case, the stress on the fourth element attached to
a node, for which the three other fault segments have ruptured, is zero by
the law
of local static mechanical equilibrium. This is consistent both with mechanical
equilibrium by definition and also with the dynamical stress drop
condition. Notice that
this is the only situation for which the condition of dynamical stress drop
is always
compatible with  static mechanical equilibrium. However, the crack model
presents a
curious and probably rather artificial behavior in this case. Remember that a
rupture cycle is characterized usually by a progressive acceleration of the
rate of small earthquakes prior to the occurrence of the run-away. Since
the run-away
spans the whole width of the lattice, and since the dynamical stress is
imposed on its
ruptured elements, this amounts to impose the total stress within the plate
equal to
zero. Previous ruptures on small faults within a cycle have produced
localized slip and
stress sources. They cannot remain in the presence of this global vanishing
of the stress
within the plate. As a consequence, we observe in fast time a cloud of
small earthquakes
accompanying the run-away, which are the ''ghosts'' of all the previous small
earthquakes. These ''ghosts''  present exactly a slip which is the opposite
of the slip
that they have develop in the foreshock phase. In other words, the
aftershocks occurring
in fast time are the exact symmetric of all the foreshocks. The difference
is that the
foreshocks are spread in time over the period of the cycle while the
aftershocks
''ghosts'' are occurring in fast time, just after the run-away. While the
details of this
behavior is clearly model specific, this phenomenon is not without
recalling field
observations that foreshocks occur usually years or even decades before a
great event,
while the huge majority of aftershocks are clustered over a few months
after the main
event. The present model does not contain however the necessary ingredients
to describe
the time delays associated with the coupling with other modes of creep or
ductile
deformations, the ductile crust and the fluid in the crust.

Ref.\cite{Bhaga} have studied the same crack model and it is instructive to
discuss how their results differs from ours in many aspects. They have only
studied the
case $\sigma_{dyn} = 0$. They have thus not found the frustration effect
discussed
above. In addition, they consider an annealed disorder, i.e. all the
threshold of the
fractures elements are re-set to new random numbers after each event. As a
consequence,
they cannot obtain earthquake localization on well-defined faults but only
observe
diffuse earthquakes. Also, this reshuffling of the disorder prevents the
synchronization to a periodic cycle and the appearence of a run-away.
Nevertheless,
they observe that the distribution of small earthquakes is a powerlaw with
$B=0.8 \pm
0.1$, not far from our estimate and that the distribution presents a peak at
large earthquakes whose energy scales with the square of the system size.
While they
interpret this as a finite size effect, we rather conclude that these large
events are
the shadows of the great run-aways in the presence of annealed noise, which
sizes are
given by that of the system. In a sense, the annealed disorder makes their
system
function permanently in our transient regime. They have only studied
disorder on the
thresholds by the method of Green functions, which is not useful
practically in the
presence of elastic disorder. The gradient conjugate method that we have
used is slower
but more general to tackle this second case. Finally, they have used
infinite system
Green functions, and have not  addressed the question of the effect of
boundaries in
finite systems. While in the statistical physics of critical phenomena, one
would like
to get results which are independent of boundaries, in the present
mechanical problem as
well as in the general mechanical case, the existence of well-defined boundary
conditions on the stress or strain fields at the border of the system is
know to control
drastically the localization of the mechanical deformation, as we have been
able to
observe.

\section{Concluding remarks}

Except for special realizations with the strongest disorder on rupture
thresholds
and elastic coefficients, we have found that the quasi-static crack model of
earthquake recurrence leads to periodic cycles, characterized by small
foreshocks
distributed according to a universal Gutenberg-Richter law with exponent
$B=1$ up to a
maximum nucleation size and a large run-away ending the cycle. This
periodic behavior
results from the strong synchronization brought by imposing a constant
dynamical
stress, corresponding to an attractive or absorptive state. The other main
result is
the discovery of a fundamental frustration resulting from an
overdetermination of the
stress field in the presence of large disorder and imposed dynamical stress
drop. The
general solution to this breakdown of self-consistency is to re-introduce a
genuine
dynamics allowing the local breakdown of static mechanical equilibrium,
associated to
the radiation of elastic waves. Our study pinpoints the fundamental role
played by
elastodynamics in repetitive crack ruptures. We thus believe that, in crack
models
(and not in dislocation models), there is no other way than incorporate the
full
elastodynamic equations to get a self-consistent solution in all situations.

We acknowledge stimulating discussions with L. Knopoff and S. Nielsen.

\pagebreak

\clearpage

{\Large \bf Figure captions}

Fig.1 :  The parameters are $\Delta \sigma = 1.9$, $\sigma_{dyn} = 0.04$, and
the constant velocity imposed at the boundary is $V=10^{-3}$.
The figure shows two maps of the accumulated slip in fault segment at two
different times
in a square system of size $L=130$ in the transient regime. We
represent those elements which have slipped at least once, the light grey
to black scale
corresponding to increasing cumulative slips. The two times of
observations correspond respectively to the first $1500$ events (top) and
$2000$ events
(bottom).  One observes a rather diffuse "damage" at
early times and localization of the deformation at long times.

\vskip 0.5cm
Fig.2 :  The parameters are $\Delta \sigma = 1.9$, $\sigma_{dyn} = 0.04$, and
the constant velocity imposed at the boundary is $V=10^{-3}$.

a) Long-time accumulated slip after localization for the same parameters as for
fig.1 but with another disorder realization. All the earthquakes occur on
this 1D fault.

b) Time dependence of the total elastic energy stored in the plate. The
periodic
behavior established after the transient up to time $50000$ is clearly visible.

c) Coding of the active fault elements as a function of time. The position
of a given
fault element is coded by a single index (denoted ''position of broken
bonds'' in the
figure) increasing from $1$ to $L^2 = 16900$. The index spans $x=1$ to $L$
at fixed $y$
for each $y=1$ to $L$. Note the periodic cycle with three phases: i)
quiescence after
the great event, ii) reactivation of the seismic activity mainly on the
main fault and
iii) increase of the foreshocks frequency in the system.

d) Gutenberg-Richter distribution of the number of events having a given
energy. The
energy of an event is defined as the difference between the total elastic
energy stored
in the system before and after the event. The crosses, plus and square symbols
corresponds to different times in the dynamics at long times, showing the
stability of the
distribution. The diamonds correspond to the distribution of small events
in a time
window in the third regime close to the run-away in the periodic regime. In
all cases, we
observe powerlaw distribution $P(E) dE \sim E^{-(1+B)}$ with a constant
exponent $B= 1
\pm 0.05$. The straight line has slope $-2$ for comparison. It is
interesting to note
that the foreshocks represented by the diamonds have the same $B$-value but
are larger on
average, signaling the nucleation of the run-away. The great earthquake is not
represented on the figure, being out of scale.

\vskip 0.5cm
Fig.3 : a) Fault localization obtained at long times for
$\Delta \sigma = 1.9$, disorder on the elastic coefficients $\Delta g = 1$,
$\sigma_{dyn} = 0.04$, and the constant velocity imposed at the boundary is
$V=10^{-3}$.
Note that the lower fault in grey becomes inactive after the transient and its
contribution to the cumulative slip vanishes at long times.

b) Gutenberg-Richter distribution of the number of events having a given
energy for
different time windows at increasing times when going from right to left.
In all cases,
we observe powerlaw distribution $P(E) dE \sim E^{-(1+B)}$ with a constant
exponent $B= 1
\pm 0.05$. The straight line has slope $-2$ for comparison. For early
times, in the
power law presents the same exponent, the events have a larger size which
finally settle
to a stationary distribution.

\vskip 0.5cm
Fig.4 : Map of a small part of the stress field
during an earthquake in fast time in a network with a power law distribution of
thresholds with $\mu_{\sigma}=0.5$ and $\sigma_{dyn} = 0.5$. The fault
segments which
have ruptured are indicated by an arrow with a thick line. These ruptured
elements all
carry a stress whose absolute value is $\sigma_{dyn} = 0.5$. The arrows
indicate the
sign of the stress: downward (resp. upward) arrows correspond to positive
(resp.
negative) stresses (recall that all the stresses are along the $Oz$ axis).
The stress carried by each fault element is indicated on the arrow. The
rupture threshold for each element is written on the other side of the bond.
A ''frustrated'' element is seen, with a stress equal to $1.5$ whose
threshold is $\sigma_c = 1.495$.


\begin{thebibliography}{99}
\bibitem{Mohr} The stress threshold rupture criterion used in our model can be
interpreted as a standard Mohr-Coulomb criterion as follows. Recall that the
Mohr-Coulomb criterion is believed to apply for
rupture in the brittle crust at depth. It states that slip occurs when the
shear stress on
a fault plane reaches a value equal to the normal stress applied on this
fault plane
times a so-called friction coefficient which is a function of roughness and
material
properties of the fault. We notice that the non-vanishing stress components
$\sigma_{xz}$ and $\sigma_{yz}$ of our antiplane model correspond to pure shear
stresses along Oz applied on the vertical plaquette borders, interpreted as
elementary
fault segments. Let us introduce a constant hydrostatic pressure in the
plane (Ox,
Oy) of the tectonic plate. This pressure creates a stress normal to the
fault segment.
According to the Mohr-Coulomb criterion, slip occurs on a given fault
segment if its
shear stress (namely either  $\sigma_{xz}$ or $\sigma_{yz}$) reaches a
constant equal
to the pressure times the coefficient of friction of this fault. This justifies
our choice of the stress threshold rupture criterion used in our model. The
heterogeneity of the thresholds $\sigma_c$ then reflects that of the friction
coefficients on different fault segments.
\bibitem{electric} Alternatively, one can interpret the map as a electric
current field,
within the usual mechanical-electrical analogy (stress $\rightarrow$ current,
displacement $\rightarrow$ voltage, elastic coefficient $\rightarrow$
conductance) which
is exact for antiplane elasticity.
\end{thebibliography}
\end{document}